\documentclass[letterpaper]{article} 
\usepackage{aaai24}  
\usepackage{times}  
\usepackage{helvet}  
\usepackage{courier}  
\usepackage[hyphens]{url}  
\usepackage{graphicx} 
\usepackage{natbib}  
\usepackage{caption} 
\usepackage{algorithm}
\usepackage{algorithmic}

\usepackage{subfigure}
\usepackage{amsmath,amssymb}
\usepackage{multirow}

\usepackage{booktabs}

\usepackage{newfloat}
\usepackage{listings}
\DeclareCaptionStyle{ruled}{labelfont=normalfont,labelsep=colon,strut=off} 
\floatstyle{ruled}
\newfloat{listing}{tb}{lst}{}
\floatname{listing}{Listing}
\title{MM-TTS: Multi-Modal Prompt Based Style Transfer for Expressive Text-to-Speech Synthesis}

\author{
    Wenhao Guan\textsuperscript{\rm 1},
    Yishuang Li\textsuperscript{\rm 2},
    Tao Li\textsuperscript{\rm 1},
    Hukai Huang\textsuperscript{\rm 1},\\
    Feng Wang\textsuperscript{\rm 1},
    Jiayan Lin\textsuperscript{\rm 1},
    Lingyan Huang\textsuperscript{\rm 1},
    Lin Li\textsuperscript{\rm 2,\rm 3$\ast$},
    Qingyang Hong\textsuperscript{\rm 1}\footnote{Corresponding author.}
}
\affiliations{
    \textsuperscript{\rm 1} School of Informatics, Xiamen University, China\\
    \textsuperscript{\rm 2} Institute of Artificial Intelligence, Xiamen University, China \\
    \textsuperscript{\rm 3} School of Electronic Science and Engineering, Xiamen University, China \\
    whguan@stu.xmu.edu.cn,\{lilin,qyhong\}@xmu.edu.cn
}

\usepackage{bibentry}

\begin{document}

\maketitle

\begin{abstract}
The style transfer task in Text-to-Speech (TTS) refers to the process of transferring style information into text content to generate corresponding speech with a specific style.  However, most existing style transfer approaches are either based on fixed emotional labels or reference speech clips, which cannot achieve flexible style transfer. 
Recently, some methods have adopted text descriptions to guide style transfer. 
In this paper, we propose a more flexible multi-modal and style controllable TTS framework named $\textit{MM-TTS}$. 
It can utilize any modality as the prompt in unified multi-modal prompt space, including reference speech, emotional facial images, and text descriptions, to control the style of the generated speech in a system. 
The challenges of modeling such a multi-modal style controllable TTS mainly lie in two aspects: 1) aligning the multi-modal information into a unified style space to enable the input of arbitrary modality as the style prompt in a single system, and 2) efficiently transferring the unified style representation into the given text content, thereby empowering the ability to generate prompt style-related voice. 
To address these problems, we propose an aligned multi-modal prompt encoder that embeds different modalities into a unified style space, supporting style transfer for different modalities. Additionally, we present a new adaptive style transfer method named Style Adaptive Convolutions (SAConv) to achieve a better style representation. Furthermore, we design a Rectified Flow based Refiner to solve the problem of over-smoothing Mel-spectrogram and generate audio of higher fidelity. Since there is no public dataset for multi-modal TTS, we construct a dataset named $\textit{MEAD-TTS}$, which is related to the field of expressive talking head. Our experiments on the MEAD-TTS dataset and out-of-domain datasets demonstrate that MM-TTS can achieve satisfactory results based on multi-modal prompts. The audio samples and constructed dataset are available at \url{https://multimodal-tts.github.io}.

\end{abstract}

\section{Introduction}

\begin{figure}[t]
\centering
\includegraphics[width=0.81\linewidth]{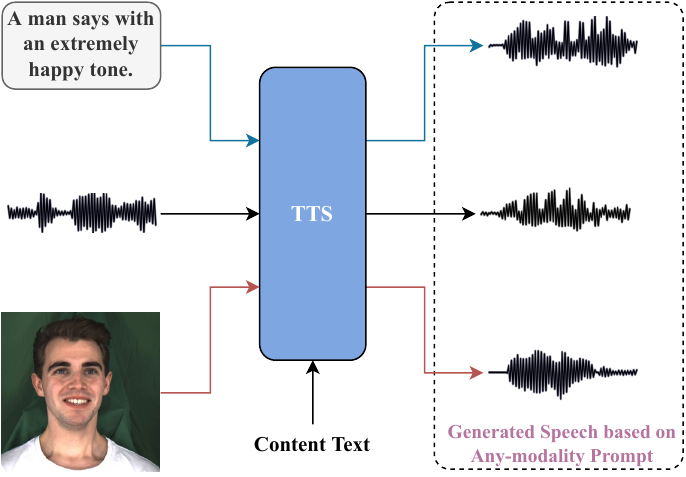}
\caption{The general style transfer framework for TTS with multi-modal prompts.}
\label{fig:task}
\end{figure}

\begin{figure*}[t]
  \centering
  \includegraphics[width=1.0\linewidth]{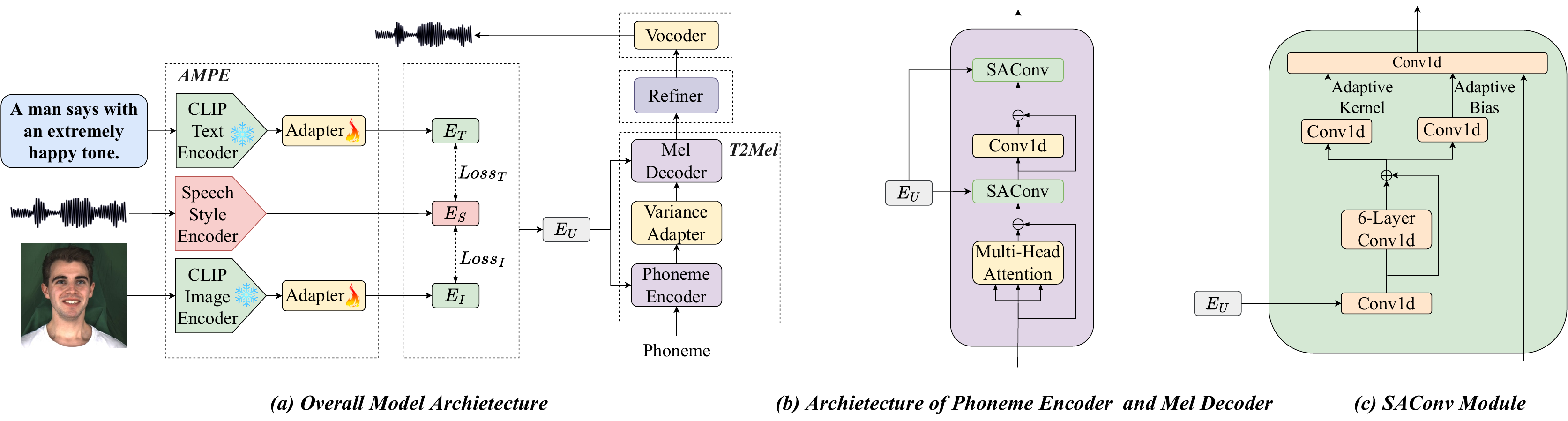}
  \caption{The model architecture of MM-TTS.  }
  \label{fig:MMTTS}
\end{figure*}

With the rapid advancement of deep learning, Text-to-Speech (TTS) has witnessed significant progress. The development of typical neural TTS systems  presents diversification. To address the challenge of slow inference speed in autoregressive models~\cite{shen2018natural,li2019neural}, non-autoregressive systems like FastSpeech~\cite{ren2019fastspeech} and  FastSpeech2~\cite{renfastspeech} have emerged as viable solutions. In order to explore the impact of different generative models on TTS systems, a series of TTS models based on generative models, including Glow-TTS~\cite{kim2020glow}, Diff-TTS~\cite{jeong2021diff}, VITS~\citep{kim2021conditional} and ProDiff~\cite{huang2022prodiff}, have been developed. Moreover, to address the mismatch issue in the Mel-spectrogram domain within the two-stage system, several end-to-end TTS models~\cite{kim2021conditional,tan2022naturalspeech,lei2023unisyn} have been proposed. Recently, novel approaches utilizing discrete tokens~\cite{du2023unicats,wang2023neural} for modeling text and speech sequences have emerged, leading to more natural and high-fidelity speech synthesis. The applications of these technologies have progressively bridged the gap between the generation result of typical TTS system and human-level speech.

The essence of TTS is a cross-modal task that maps text content to corresponding speech. However, simply generating speech corresponding to the text content is not enough. We also hope that the generated speech has rich paralinguistic information such as speaker identity, language, emotion and prosody.
A lot of expansion works are proposed to achieve specific purposes. For example, multi-speaker TTS~\citep{casanova2021sc,chen20r_interspeech}, multilingual TTS~\citep{li2021light}, style transfer~\citep{huang2022generspeech,guan23_interspeech}. 
We divide the style transfer tasks for TTS into four main categories: 
emotion label based style transfer, reference speech based style transfer, face based style transfer and text description based style transfer.
Emotion label based style transfer~\citep{lee2017emotional,lorenzo2018investigating} uses several predefined kinds of emotional category labels to represent the style.
Reference speech based style transfer~\citep{min2021meta,huang2022generspeech,guan23_interspeech} aims at extracting the style of given speech segment and transferring the style to generate style-related speech for any text content. 
Face based Style Transfer~\citep{goto2020face2speech,wang2022residual} mainly focuses on extracting the speaker identity of a given face image and then it can be applied in multi-speaker TTS. 
Text description based style transfer~\citep{guo2023prompttts,yang2023instructtts}  utilizes natural language prompts to generate specific styles of speech. 
Although these systems can achieve good results, they are not flexible enough. In this paper, we propose a more flexible style transfer framework, which is not constrained by a single modality prompt and can input any modality as the prompt during inference stage.

To achieve more flexible style controllability, we propose a general TTS framework with multi-modal prompts as shown in Figure \ref{fig:task}.
We can input any modality into this system and subsequently provide any textual input as the content text. The system is capable of generating speech that is stylistically related to the corresponding modality.
Specifically,  we propose a novel TTS method, $\textit{MM-TTS}$.
Figure \ref{fig:MMTTS} illustrates the architecture of MM-TTS for controllable style transfer with multi-modal prompts.
To achieve alignment among multi-modal prompt features, we introduce an Aligned Multi-modal Prompt Encoder (AMPE) based on the semantic understanding capability of Contrastive Language-Image Pretraining (CLIP)~\citep{radford2021learning} to unify the text, speech and image prompt modality into the unified style space, thus supporting flexible style control via multi-modal inputs. 
For our text-to-mel model, we propose a Style Adaptive Convolutions (SAConv) module to extract more local details of style information, which also facilitates the style transfer ability.
To overcome the over-smoothness problem produced by the text-to-mel model~\citep{ren2022revisiting}, we design a novel module called Reflow Refiner, which is based on the Rectified Flow~\citep{liu2022flow} to obtain the Mel-spectrogram closer to the real domain.
Finally the generated Mel-spectrogram is converted to waveform by a pretrained vocoder.

However, there is no public dataset for the multi-modal TTS,
in this paper, we construct a dataset that consists of text, face and speech prompts for expressive and controllable TTS  based on the emotional talking head video dataset MEAD \cite{wang2020mead}. We call the newly constructed multi-modal expressive TTS dataset as MEAD-TTS.

In summary, the main contributions of this work are as follows:
\begin{itemize}
	\item We propose a novel method MM-TTS to endow the TTS with multi-modal capabilities. To our knowledge, this is the first multi-modal prompt based style transfer framework for TTS. Abundant experiments are conducted to demonstrate the superiority of our method on objective and subjective evaluations.
    \item To unify the multi-modal prompt style space, we propose a novel AMPE, allowing flexible style control guided by prompts of different modalities for TTS. Additionally, we introduce a novel module, SAConv, to transfer the style information extracted by the AMPE into any content text to generate style-related speech more effectively.
    \item A novel Rectified Flow based Refiner is designed to refine the previously generated Mel-spectrograms, thus resolve the over-smoothness problem and ultimately get Mel-spectrograms closer to the real data domain.
    \item We construct a dataset MEAD-TTS for multi-modal TTS,  providing the reference dataset for future works.
\end{itemize}

\section{Related Work}
\subsection{Reference Speech Based Style Transfer}
Reference speech based style transfer is the most popular method for style transfer in TTS, because the speech clips are easy to obtain and contain rich information.
Global style token (GST)~\cite{wang2018style} designs a style token layer and a  reference encoder to explore the expressiveness of TTS systems unsupervisedly. VAE-Tacotron~\cite{zhang2019learning} learns the style representation through VAE ~\cite{kingma2013auto}.
Subsequently, significant efforts have focused on designing robust style encoders for reference speech based style transfer tasks. Some works, based on autoregressive models, have achieved fine-grained style representations through the design of multi-level or multi-scale style modeling methods ~\cite{sun2020generating,li21r_interspeech}. However, to overcome the issue of low decoding speed associated with autoregressive models, non-autoregressive architectures have been adopted in some methods. Meta-StyleSpeech~\cite{min2021meta} adopts the base architecture upon FastSpeech2~\citep{renfastspeech}, applying style adaptive layer norm and meta-learning algorithm to effectively synthesize style-transferred speech. GenerSpeech~\cite{huang2022generspeech} proposes a multi-level style adaptor and a generalizable content adaptor to efficiently model the style information. 
IST-TTS~\cite{guan23_interspeech} designs a novel TTS system that can perform style transfer with interpretability and high fidelity based on controllable variational autoencoder ~\cite{shao2020controlvae} and diffusion models. 
In this work, we propose a more effective SAConv module to transfer extracted style information and thus we obtain the highly style-related speech.

\subsection{Face Based Style Transfer}
We can imagine speakers' voice characteristics from their faces. Due to the consideration of this phenomenon, there are some works that utilize facial features to generate speech that matches the speaker's characteristics. Face2Speech \cite{goto2020face2speech} uses a face image to control the voice characteristics of the synthesized speech by training a face encoder and a speaker encoder separately and finally designs a loss function to make their embeddings closer. In FR-PSS \cite{wang2022residual}, based on Face2Speech, a residual-guided strategy is designed by incorporating a prior speech feature to make the network capture representative face features. The works mentioned above train the face encoder to share a joint embedding space with the speech encoder, independently from the main TTS model. FaceTTS \cite{lee2023imaginary} designs a multi-speaker TTS model by training a robust cross-modal representation of speaking style, where speaking styles are conditioned on face attributes. But existing works usually focus on extracting the speaker identity of a given face image for multi-speaker TTS, which is limited in style-rich TTS systems. In this paper, we focus on modeling the style attributes of facial images such as gender and emotion.

\subsection{Text Description Based Style Transfer} 
Due to the emergence of large language models and text based image generation, using a text description as prompt to guide the generation of texts or images has drawn wide attention recently. In the field of TTS, several recent works \cite{kim21n_interspeech,guo2023prompttts,yang2023instructtts,liu23t_interspeech} have emerged to utilize text descriptions to guide speech synthesis.
PromptTTS \cite{guo2023prompttts} designs a style encoder, which maps a style prompt to a semantic space to extract the style representation, to guide the content encoder and the speech decoder.
InstructTTS \cite{yang2023instructtts} introduces a three-stage training strategy to obtain a robust embedding model, which can effectively capture semantic information from the style prompts. It proposes to model acoustic features in discrete latent space and utilizes discrete diffusion model to generate discrete acoustic features.
PromptStyle \cite{liu23t_interspeech} proposes a two-stage TTS approach for cross-speaker style transfer with natural language descriptions based on VITS \cite{kim2021conditional}. 
In this paper, we utilize the strong semantic capability of CLIP to obtain the style representations of text descriptions.

\section{MM-TTS}

\subsection{Overview}
MM-TTS is considered a multi-modal prompt based style transfer framework for expressive TTS. 
In order to achieve highly flexible multi-modal style transfer for TTS, we design several modules to achieve the unified style space and the efficient adaptive style transfer respectively.

As shown in Figure \ref{fig:MMTTS}, the proposed framework consists of four modules: 1) an aligned multi-modal prompt encoder (AMPE), which includes a CLIP based text encoder, a CLIP based image encoder, a Speech style encoder and two Adapter modules, to align the multi-modal information into a unified style space; 
2) a Text-to-Mel model (T2Mel) that maps the given text content to corresponding speech, where we introduce Style Adaptive Convolutions (SAConv) to efficiently transfer the unified style representation to generate the style related speech. We adopt the StyleSpeech proposed in~\citep{min2021meta}, which is based on FastSpeech2~\citep{renfastspeech}, as the T2Mel backbone;
3) a Refiner to refine the Mel-spectrograms and get more realistic Mel-spectrograms.
4) a vocoder to convert the Mel-spectrogram to waveform.

\subsection{Aligned Multi-Modal Prompt Encoder}
To get a unified multi-modal prompt based style space, we introduce an Aligned Multi-modal Prompt Encoder (AMPE) based on CLIP to unify the text, speech and image prompt modality into the unified style space, thus supporting flexible style control via multi-modal inputs. 

In practice, in the training phase, we first input prompts of three modalities (text, speech, face) into the AMPE module. Specifically, the fixed CLIP based text and image encoders followed by learned adapter modules are leveraged to extract corresponding text prompt embedding $\boldsymbol E_{T}$ and face image prompt embedding $\boldsymbol E_{I}$, and the speech style encoder is learned to extract the speech prompt embedding $\boldsymbol E_{S}$. 
The unified style embedding $\boldsymbol E_{U}$ is guided by several MSE loss functions, which is computed as follows:
\begin{equation}
\begin{aligned}
L_{AMPE}&=Loss_{I}+Loss_{T}\\
&=MSE(\boldsymbol E_{I},\boldsymbol E_{S})+MSE(\boldsymbol E_{T},\boldsymbol E_{S}),
\end{aligned}
\label{eq:clip_loss}
\end{equation}

The unified style embedding $\boldsymbol E_{U}$ is different at training phase and inference phase:
\begin{equation}
\begin{aligned}
&\boldsymbol E_{U} = \boldsymbol E_{S}, \text{at~the~training~phase}  \\
&\boldsymbol E_{U} \in \left\{\boldsymbol E_{S},\boldsymbol E_{I},\boldsymbol E_{T}\right\}, \text{at~the~inference~phase}
\end{aligned}
\label{eq:clip_loss}
\end{equation}

In the inference phase, we can employ arbitrary embedding of different modalities as the $\boldsymbol E_{U}$.

\begin{figure}[t]
  \centering
  \includegraphics[width=0.77\linewidth]{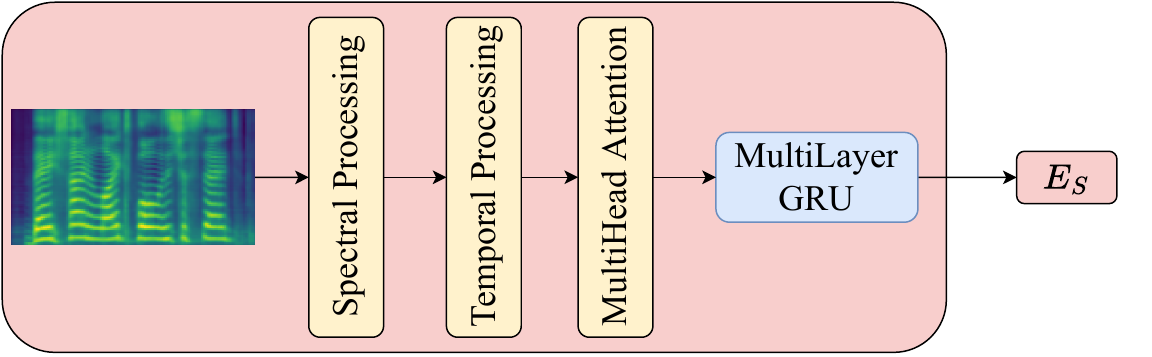}
  \caption{The architecture of speech style encoder in MM-TTS. }
  \label{fig:SAConv}
\end{figure}

For the speech style encoder in the AMPE module, 
as shown in Figure \ref{fig:SAConv}, we design the  speech style encoder including four components. The initial three components, namely Spectral Processing, Temporal Processing, and Multi-head Attention, are analogous to those employed in Meta-StyleSpeech~\citep{min2021meta}. Additionally, we introduce a fourth component, a Multi-layer GRU~\citep{chung2014empirical}, to capture richer style information. Finally, we get an informative multi-channel vector.
The adapter modules in AMPE are simply comprised of two fully connected layers.

\subsection{Style Adaptive Convolutions}
In previous works~\citep{casanova2021sc,guan23_interspeech}, the extracted style information is usually directly fed into the generator through concatenation or summation. 
Meta-StyleSpeech~\citep{min2021meta} proposes Style Adaptive Layer Normalization (SALN) to transfer the statistical properties of reference style features to a given content text, and thus gets style transferred speech.
Inspired from Adaptive Convolutions ~\citep{chandran2021adaptive} for image style transfer, we propose Style Adaptive Convolutions (SAConv) to transfer the reference style features more precisely.

For the architecture, as illustrated in Figure \ref{fig:MMTTS} (c), the SAConv receives the style embedding $\boldsymbol E_{U}$ and predicts the kernel and bias via kernel prediction networks, then the predicted kernel and bias are used for context feature to get prompt style transferred speech.
Specifically, the kernel prediction network comprises of an input convolution, a residual convolution module with 6 convolution layers, a kernel convolution and a bias convolution to predict adaptive kernel and bias respectively.

Given a context feature $\boldsymbol x$ and the desired style embedding $\boldsymbol E_{U}$, the normalized context feature $\boldsymbol x_{norm}$ is derived as follows:
\begin{equation}
\boldsymbol x_{norm}=\frac{\boldsymbol x - \mu_{x}}{\sigma_{x}}
\end{equation}
where $\mu_{x}$ and $\sigma_{x}$ are the mean and standard deviation of the input context feature $\boldsymbol x$. 

Then, we obtain the predicted kernel $\boldsymbol{conv}_{kernel}$ and bias $\boldsymbol{conv}_{bias}$ of the given style feature $\boldsymbol E_{U}$ by the kernel prediction network.
And then the SAConv is computed utilizing predicted kernel $\boldsymbol{conv}_{kernel}$ and bias $\boldsymbol{conv}_{bias}$ as follows:
\begin{equation}
\begin{aligned}
SAConv(\boldsymbol x, \boldsymbol{conv}_{kernel}, \boldsymbol{conv}_{bias})\\
=\boldsymbol{conv}_{kernel}(\boldsymbol x_{norm})+\boldsymbol{conv}_{bias}
\end{aligned}
\end{equation}

While the gain and bias of SALN are fully connected layers, which only transfer global statistics of style features, our SAConv aims at predicting convolution kernels and biases according to the given style embedding, to transfer the style features more precisely. The detailed model information is in Appendix C.

\begin{figure*}[!t]
  \centering
  \includegraphics[width=0.77\linewidth]{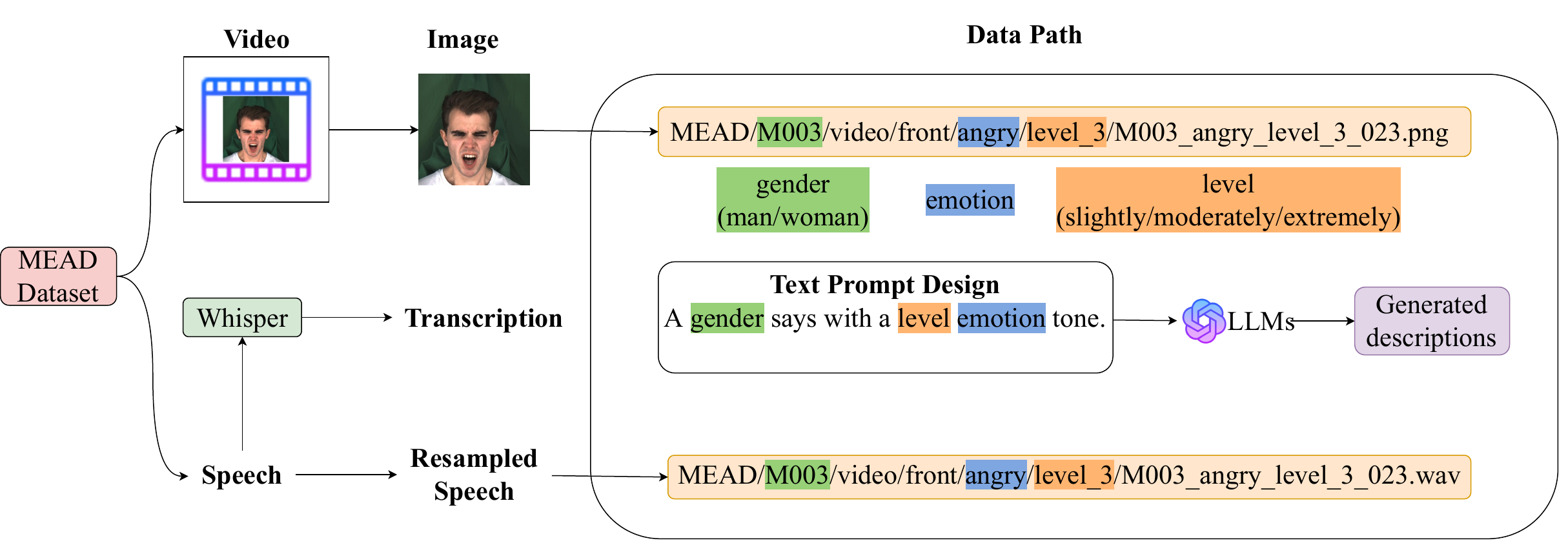}
  \caption{The data preprocessing pipeline for MM-TTS. }
  \label{fig:data_pipeline}
\end{figure*}

\subsection{Rectified Flow Based Refiner}

In order to overcome the over-smoothing problem and maintain the effectiveness of our SAConv module to the greatest extent, we propose a novel rectified flow~\citep{liu2022flow} based refiner (Reflow Refiner) through two-stage TTS training pipeline like the way in IST-TTS~\citep{guan23_interspeech}, which can yield high-quality results for Mel-spectrogram generation when simulated with a very few number of Euler steps. 
The Reflow Refiner is an Ordinary Differential Equation (ODE) model that transports distribution $\pi_{0} $ to $\pi_{1}$ by straight line paths as much as possible, where $\pi_{0}$ is the standard gaussian distribution and $\pi_{1}$ is the ground truth distribution.
Given empirical observations of $X_{0}\sim\pi_{0}, X_{1}\sim\pi_{1}$, the Reflow Refiner induced from $(X_{0},X_{1})$ is an ODE conditioned on generated Mel-spectrogram $c_{mel}$ in the first training stage with respect to time $t \in [0,1]$, 
\begin{equation}
dX_{t}=v(X_{t},t,c_{mel})dt,
\end{equation}
which converts $X_{0}$ from $\pi_{0}$ to a $X_{1}$ from $\pi_{1}$. The drift force $v:\mathbb{R}^{d} \rightarrow \mathbb{R}^{d}$ is set to drive the flow to follow the direction $(X_{1}-X_{0})$ of the linear path pointing from $X_{0}$ to $X_{1}$ as much as possible, by solving a simple least squares regression problem, the loss $L_{Refiner}$ is as follows:
\begin{equation}
\begin{aligned}
L_{Reflow}=\int_{0}^{1} \mathbb{E}[\parallel(X_{1}-X_{0})-v(X_{t},t,c_{mel})\parallel ^{2}]dt,  \\
\end{aligned}
\end{equation}
where $X_{t}=tX_{1}+(1-t)X_{0}$, $X_{t}$ is the linear interpolation of $X_{0}$ and $X_{1}$. 

During inference phase, we get $v$ and we solve the ODE starting from $X_{0}\sim\pi_{0}$ to transfer $\pi_{0}$ to $\pi_{1}$ conditioned on $c_{mel}$. Our MM-TTS utilizes the ODE RK45 sampler for inference.

With the training of the Reflow Refiner, it can make the paths between  $\pi_{0}$ and  $\pi_{1}$ tending to straight lines. 
As we expected, perfectly straight paths can be simulated exactly with a single Euler step, which is a one-step model with high quality and fast inference speed. 
We put more detailed information about the Reflow Refiner module in Appendix C.

\subsection{Objectives}
Our training pipeline has two stages: text-to-mel stage and refiner stage. Combining the above AMPE, SAConv and text-to-mel model, we optimize the  model in the first stage by minimizing the loss function as:
\begin{equation}
L_{1s}=L_{Mel}+L_{Var}+L_{AMPE},
\end{equation}
where $L_{Mel}$ denotes the MAE loss between the ground-truth and Mel-spectrogram that generated by the Mel Decoder in MM-TTS. 
$L_{Var}$ denotes the duration, pitch and energy reconstruction loss.

In the second stage for Reflow Refiner training, we optimize the Reflow Refiner model by minimizing the loss function as:
\begin{equation}
L_{2s}=L_{Reflow}
\end{equation}

\section{MEAD-TTS Dataset}
To the best of our knowledge, there is no public expressive dataset for TTS with multi-modal prompts, so we construct a dataset to endow TTS with multi-modal prompt capability based on MEAD~\cite{wang2020mead}, which is originally a dataset of expressive talking head generation.
The data preprocessing pipeline is illustrated as Figure \ref{fig:data_pipeline}.
To get transcriptions of corresponding speech clips, we utilize the whisper tool~\cite{radford2023robust}.  Specifically, we designate the language as English, and using template matching method to match transcription texts given in MEAD. To get face images corresponding to the given speech clip, we randomly select two video frames to represent the corresponding face images. 
Moreover, we design a text prompt template to construct text description prompts corresponding to different speech clips. 
The text prompt  template has three main style variables: gender, emotion and emotion level, where the emotion has eight types (neutral, angry, contempt, disgusted, fear, happy, sad and surprised).
So the style of this work refers to the summation of the three style variables (gender, emotion, emotion level).
We get different text prompt descriptions by extracting the three variables from different data paths, such as ``A $<$gender$>$ says with a $<$emotion  level$>$ $<$emotion$>$ tone". 
Finally, we utilize the remarkable generative capability of current LLMs (i.e. GPT-3.5-TURBO) to generate the text descriptions having the same meaning as previous templated prompts. When constructing the paired dataset, we center around a certain speech clip, randomly select a corresponding face image and the corresponding generated text description by LLMs as a data pair.

Through processing on MEAD dataset, the MEAD-TTS dataset consists of 31055 pairs of (speech,~face~image,~text~description) data with a total duration of approximately 36 hours for speech modality. In practice, the speech clips are resampled to 16kHz. The more details of the MEAD-TTS dataset is in Appendix A.

\section{Experimental Setup}
\subsection{Datasets}

Because the text transcriptions of MEAD-TTS dataset are limited, we first pretrain the model on LJSpeech~\citep{ito2017lj} to increase the number of text transcription data and strengthen the output diversity of TTS, where the model only receives the speech modality as prompt when training on LJSpeech. 

In order to comprehensively evaluate the effectiveness of the proposed method, we conduct experiments in intra-domain and out-of-domain datasets respectively.
We utilize the MEAD-TTS dataset for intra-domain evaluation.
And the LibriTTS~\citep{zen2019} transcriptions and speech clips are utilized for out-of-domain evaluation of reference speech based style transfer. 
For face based style transfer, we use the face images in Oulu-CASIA~\citep{zhao2011facial} dataset and transcriptions in LibriTTS for out-of-domain evaluation. 
For text description based style transfer, we use LibriTTS  transcriptions for out-of-domain evaluations. 
The more details of the used datasets are in Appendix B.
\begin{table*}[t]
\centering
\resizebox{136mm}{19mm}{
\begin{tabular}{l|cccc|cccc}
\toprule
\multirow{2}{*}{\textbf{Method}} & \multicolumn{4}{c|}{\textbf{Intra-domain}} &  \multicolumn{4}{c}{\textbf{Out-of-domain}}\\ \rule{0pt}{10pt} 
 & MOS & SMOS & SECS & MCD  & MOS & SMOS & SECS & MCD  \\ 
 \midrule
 GT  & 4.51 $\pm$ 0.04 & -  & - &  - & 4.72 $\pm$ 0.05 &  - & - &  -   \\ 
 GT (Mel+Voc) & 4.49 $\pm$ 0.08 & 4.81 $\pm$ 0.06 & 0.989 & 1.02  & 4.69 $\pm$ 0.08 &  4.79 $\pm$ 0.07 & 0.935 & 1.05  \\ \midrule  
 MS-FastSpeech2 & 3.91 $\pm$ 0.05 & 3.96 $\pm$ 0.05 & 0.924 & 3.90 & 3.43 $\pm$ 0.04 &  3.34 $\pm$ 0.05  & 0.682 & 7.74 \\  
 GenerSpeech & 4.16 $\pm$ 0.04 & 4.41 $\pm$ 0.04 & 0.904 & 3.76 & 3.87 $\pm$ 0.08 &  3.89 $\pm$ 0.04 & \textbf{0.732} & 7.25  \\  
 IST-TTS & 4.19 $\pm$ 0.06 & 4.52 $\pm$ 0.04 & 0.927 & 3.72 & 3.81 $\pm$ 0.08 &  3.83 $\pm$ 0.04 & 0.696 & 7.31  \\ 
 StyleSpeech & 4.09 $\pm$ 0.03 & 4.51 $\pm$ 0.03 & 0.945 & 3.26 & 3.69 $\pm$ 0.05 &  3.84 $\pm$ 0.03 & 0.716 & 6.91  \\ \midrule  
MM-StyleSpeech (ours) & 4.11 $\pm$ 0.07 & 4.49 $\pm$ 0.06 & 0.947 & 3.24 & 3.72 $\pm$ 0.09 &  3.82 $\pm$ 0.07 & 0.713 & 6.93  \\ 
MM-TTS (ours)  & \textbf{4.36 $\pm$ 0.06} & \textbf{4.61 $\pm$ 0.07} & \textbf{0.956} & \textbf{3.17} & \textbf{3.95 $\pm$ 0.08} &  \textbf{4.03 $\pm$ 0.08} & 0.728 & \textbf{6.69}\\
\bottomrule
\end{tabular}}
\caption{The performance comparison of parallel reference speech based style transfer.}
\label{tab:speech_prompt_subj}
\end{table*}

\begin{table}[h]
        \centering
        \vspace{2mm}
        \resizebox{\linewidth}{!}{
        \begin{tabular}{l|c|ccc|c|ccc}
        \toprule
        \multirow{3}{*}{\bfseries{Baseline}} & \multicolumn{4}{c|}{\textbf{Intra-domain}} & \multicolumn{4}{c}{\textbf{Out-of-domain}} \\ \cline{2-9}
        \rule{0pt}{10pt}& \multirow{2}{*}{7-point score} & \multicolumn{3}{c|}{Perference} & \multirow{2}{*}{7-point score} & \multicolumn{3}{c}{Perference} \\ 
        \rule{0pt}{10pt}      &    & C & E & O     &   & C   & E  & O  \\
        \midrule
        MS-FastSpeech2              &  1.81 $\pm$ 0.11 & 16\%  & 28\% &  56\%   & 1.62 $\pm$ 0.09   & 7\%  & 20\% &  73\% \\
        GenerSpeech      &  1.15 $\pm$ 0.12  & 28\%  & 30\% &  42\%   & 1.29 $\pm$ 0.05    & 33\%  & 21\% &  46\% \\
        IST-TTS&  1.12 $\pm$ 0.09  & 30\%  & 25\% &  45\%   & 1.27 $\pm$ 0.06    & 36\%  & 18\% &  46\% \\
        StyleSpeech       &  1.27 $\pm$ 0.07  & 27\%  & 26\% &  47\%   & 1.31 $\pm$ 0.10   & 30\%  & 20\% &  50\% \\
        MM-StyleSpeech    &  1.01 $\pm$ 0.08  & 33\%  & 27\% & 40\%   & 1.12 $\pm$ 0.07   & 26\%  & 35\% &  39\% \\
        \bottomrule 
        \end{tabular}}
        \caption{The AXY preference test results for   non-parallel reference speech based style transfer. C, E, O denote the preference rate for compared model, equivalent and our model respectively.}
    \label{table:npst}
\end{table}

\subsection{Evaluation Metrics}
We evaluate the style transfer quality and similarity by objective metrics and subjective evaluations. The objective metrics include speaker embedding cosine similarity (SECS) and mel cepstral distortion (MCD) metric. 
The SECS calculates the cosine similarity between the embeddings of two audios extracted from the speaker encoder. It ranges from -1 to 1, and a larger value indicates a stronger similarity. The MCD evaluates the spectral distance between the reference and synthesized Mel-spectrum features.
As for subjective evaluations, we conduct 5-scale Mean Opinion Score (MOS) and Similarity Mean Opinion Score (SMOS) test between MM-TTS and the baseline models.  The score of MOS test ranges from 1 to 5 with an interval of 0.5, in which 1 means very bad and 5 means excellent. Both MOS and SMOS are reported with 95\% confidence interval.
We generate 50 speech samples for each model, which are listened by 20 listeners for subjective evaluations.
We additionally evaluate the emotion classification accuracy and gender classification accuracy for face prompt and text prompt based style transfer.

Note that in our experiments, the SECS scores are calculated using the speaker encoder in Resemblyzer\footnote{\url{https://github.com/resemble-ai/Resemblyzer}.}. For emotion and gender classification, we  utilize a pretrained wav2vec~2.0~\citep{baevski2020wav2vec} model as feature extractor, and then add two fully connected layers and one softmax layer for training.

\subsection{Training Setting and Baseline Model}
The proposed MM-TTS was trained for 200K iterations using Adam optimizer~\citep{kingma2014adam} on a single NVIDIA GeForce RTX 2080Ti GPU for both the first text-to-mel stage and second refiner stage training pipeline. Additionally, we utilize a pretrained HiFi-GAN~\citep{kong2020hifi} as the neural vocoder to convert generated Mel-spectrogram to waveform.

We compare our system with other methods from three aspects:
reference speech based style transfer, face based style transfer and text description based style transfer.

As there is no related works for multi-modal TTS, 
we utilize StyleSpeech proposed in~\citep{min2021meta} endowed with multi-modal prompts, which is named as MM-StyleSpeech, as our baseline system. MM-StyleSpeech uses the similar AMPE module as MM-TTS and other modules are the same as original Stylespeech. 

For reference speech based style transfer, we conduct experiments on the following systems :
1) GT: This is the ground-truth recording; 
2) GT (Mel + Voc): This is the speech synthesized using pretrained HiFi-GAN vocoder for GT Mel-spectrogram; 
3) MS-FastSpeech2: This is a multi-speaker FastSpeech2~\citep{renfastspeech} system with x-vector speaker embedding to the encoder output and the decoder input; 
4) GenerSpeech~\citep{huang2022generspeech}: This is a style transfer model for TTS with a multi-level style adapter and a generalizable content adapter; 
5) IST-TTS~\citep{guan23_interspeech}: This is a style transfer model for TTS with controllable VAE and diffusion models; 
6) StyleSpeech~\citep{min2021meta}: This is a style transfer model for TTS proposed in~\citep{min2021meta} with Style Adaptive Layer Normalization; 
7) MM-StyleSpeech: This is our proposed baseline model for multi-modal TTS based on StyleSpeech;
8) MM-TTS: This is our proposed model for multi-modal TTS.

Previous face based style transfer methods focus on modelling the speaker identity of face images and previous text description based style transfer methods construct the models based on different self-built datasets, so we only compare with the baseline MM-StyleSpeech.

\begin{table}[t]
\centering
\resizebox{\linewidth}{!}{
\begin{tabular}{l|ccc|ccc}
\toprule
\multirow{2}{*}{\textbf{Method}} & \multicolumn{3}{c|}{\textbf{Intra-domain}} &  \multicolumn{3}{c}{\textbf{Out-of-domain}}\\ \rule{0pt}{10pt} 
 & MOS & $Acc_{emo}$ & $Acc_{gen}$  & MOS & $Acc_{emo}$ & $Acc_{gen}$   \\ \midrule  
 GT  & 4.51 $\pm$ 0.04 & 0.826  & 1.0  &  - & -  & -  \\ 
 GT (Mel+Voc) & 4.49 $\pm$ 0.08 & 0.813  & 1.0  &  - & -  & -   \\ 
 \midrule
MM-StyleSpeech  & 3.98 $\pm$ 0.05 & 0.651  & 0.996  &  3.56 $\pm$ 0.07 & 0.249  &  0.686  \\  
MM-TTS        & \textbf{4.11 $\pm$ 0.06} & \textbf{0.659}  & \textbf{0.997} &  \textbf{3.77 $\pm$ 0.08}  & \textbf{0.261}  & \textbf{0.833} \\ \bottomrule
\end{tabular}}
\caption{The performance comparison of face based style transfer.}
\label{tab:face_prompt}
\end{table}

\section{Experiment Results and Analysis}
For different modality prompts, we conduct subjective and objective evaluations respectively.

\subsection{Results on Reference Speech Based Style Transfer}
For reference speech based style transfer, we classify the experiments into two categories according to the content consistency between the reference and generated speech clips: Parallel Style Transfer (PST) and Non-Parallel Style Transfer (NPST).
Table \ref{tab:speech_prompt_subj} shows the subjective and objective results of PST including MOS, SMOS, SECS and MCD. Given the reference speech, our MM-TTS achieves better results on both audio naturalness and style similarity on MEAD-TTS dataset and out-of-domain datasets.
Our method achieves the highest SECS value and the lowest MCD value except the SECS value in out-of-domain dataset.
The results indicate that our method is more effective for extracting and transferring the style of reference speech.

For NPST, we select 100 samples from MEAD-TTS and LibriTTS testing sets for intra-domain and out-of-domain evaluation respectively. An AXY test used in ~\citep{huang2022generspeech} is conducted, the range of 7-point score is from -3 to 3, 0 represents
“Both are about the same distance”.
As shown in Table \ref{table:npst}, the results indicate that listeners prefer MM-TTS synthesis against the compared models. The proposed SAConv significantly improves the style extraction ability, allowing an arbitrary reference sample to guide the stylistic synthesis of arbitrary content text. We put the visualization results about PST and NPST in Appendix D.1.

\subsection{Results on Face Based Style Transfer}

Table \ref{tab:face_prompt} presents the experimental results of face-based style transfer, encompassing MOS  and classification accuracy for emotion and gender. In the context of face prompts, our MM-TTS model surpasses the baseline model in terms of audio naturalness and classification performance on the MEAD-TTS dataset as well as out-of-domain datasets.
These results highlight the effectiveness of our method in accurately capturing and extracting style attributes from facial images, indicating its good capability in incorporating facial styles into synthesized speech.

\subsection{Results on Text Description Based Style Transfer}
Table \ref{tab:text_prompt} displays the experimental results of text description based style transfer. By utilizing the text prompts, our MM-TTS  achieves superior scores in terms of audio naturalness and classification accuracy for emotion and gender.
The results indicate that our method has good ability in modeling and incorporating the style of natural language descriptions.

\begin{table}[h]
\centering
\resizebox{\linewidth}{!}{
\begin{tabular}{l|ccc|ccc}
\toprule
\multirow{2}{*}{\textbf{Method}} & \multicolumn{3}{c|}{\textbf{Intra-domain}} &  \multicolumn{3}{c}{\textbf{Out-of-domain}}\\ \rule{0pt}{10pt} 
 & MOS & $Acc_{emo}$ & $Acc_{gen}$  & MOS & $Acc_{emo}$ & $Acc_{gen}$   \\ \midrule  
 GT  & 4.51 $\pm$ 0.04 & 0.826  & 1.0  &  - & -  & -  \\ 
 GT (Mel+Voc) & 4.49 $\pm$ 0.08 & 0.813  & 1.0  &  - & -  & -   \\ \midrule 
MM-StyleSpeech  & 3.99 $\pm$ 0.04 & 0.635  & 1.0  &  3.73 $\pm$ 0.08 & 0.289  &  0.989 \\   
MM-TTS        & \textbf{4.19 $\pm$ 0.07} & \textbf{0.636}  & \textbf{1.0} &  \textbf{3.93 $\pm$ 0.08}  & \textbf{0.318}  & \textbf{0.996} \\  
\bottomrule
\end{tabular}}
\caption{The performance comparison of text description based style transfer.}
\label{tab:text_prompt}
\end{table}

\begin{table}[!h]
\centering
\resizebox{\linewidth}{!}{
\begin{tabular}{l|cc|cc}
\toprule
\multirow{2}{*}{\textbf{Method}} & \multicolumn{2}{c|}{\textbf{Intra-domain}} &  \multicolumn{2}{c}{\textbf{Out-of-domain}}\\ \rule{0pt}{10pt} 
 & MOS & SMOS & MOS  & SMOS  \\ \midrule   
MM-TTS   & \textbf{4.36 $\pm$ 0.06} & \textbf{4.61 $\pm$ 0.07}  & \textbf{3.95 $\pm$ 0.08} &  \textbf{4.03 $\pm$ 0.08}  \\ \midrule 
w/o SAConv  & 4.22 $\pm$ 0.07 & 4.55 $\pm$ 0.05 & 3.85 $\pm$ 0.09 &  3.81 $\pm$ 0.07     \\ 
w/o Reflow Refiner  & 4.17 $\pm$ 0.05 & 4.57 $\pm$ 0.06  & 3.79 $\pm$ 0.04 &  3.99 $\pm$ 0.05   \\ 
w/o Pretrain  & 4.32 $\pm$ 0.03 & 4.59 $\pm$ 0.04 & 3.61 $\pm$ 0.07 &  3.63 $\pm$ 0.06     \\  
\bottomrule
\end{tabular}}
\caption{The performance of ablation studies in MM-TTS.}
\label{tab:ablation}
\end{table}

\subsection{Ablation Studies}
The MOS and SMOS results of ablation studies are illustrated in Table \ref{tab:ablation}, where ``w/o SAConv'' denotes substituting the SAConv module with SALN module in StyleSpeech, ``w/o Reflow Refiner'' denotes removing the second refiner training stage  and  ``w/o Pretrain'' denotes only using MEAD-TTS dataset for training without LJSpeech pretraining.
It demonstrates that 1) substituting the SAConv module with SALN module results in a significant drop on style similarity, which demonstrates that SAConv can extract more effective style representations than SALN;
2) removing the Reflow Refiner leads to a decline in audio naturalness, which indicates that the Reflow Refiner mainly contributes to the fidelity maintaining;
3) without the pretraining on LJSpeech leads to a significant drop on naturalness and similarity especially on out-of-domain evaluations.
We put more visualization results about ablation studies in Appendix D.2.

\begin{table}[!h]
\centering
\resizebox{\linewidth}{!}{
\begin{tabular}{l|cc|cc}
\toprule
\multirow{2}{*}{\textbf{Method}} & \multicolumn{2}{c|}{\textbf{Intra-domain}} &  \multicolumn{2}{c}{\textbf{Out-of-domain}}\\ \rule{0pt}{10pt} 
 & MOS & SMOS & MOS  & SMOS  \\ \midrule  
DDPM (1000 steps)    & 4.35 $\pm$ 0.05 & 4.61 $\pm$ 0.07  & \textbf{3.96 $\pm$ 0.04} &  4.03 $\pm$ 0.03   \\ 
DDPM (100 steps)    & 4.31 $\pm$ 0.04 & 4.59 $\pm$ 0.03  & 3.89 $\pm$ 0.05 &  4.02 $\pm$ 0.05   \\   \midrule 
Reflow (1 step)  & 4.32 $\pm$ 0.05 & 4.56 $\pm$ 0.06  & 3.91 $\pm$ 0.04 &  4.01 $\pm$ 0.06   \\ 
Reflow (MM-TTS)   & \textbf{4.36 $\pm$ 0.06} & \textbf{4.61 $\pm$ 0.07}  & 3.95 $\pm$ 0.08 &  \textbf{4.03 $\pm$ 0.02}  \\
\bottomrule
\end{tabular}}
\caption{The performance of utilizing different refiners for the proposed MM-TTS in speech based style transfer task.}
\label{tab:refiners}
\end{table}

\subsection{Refiner Evaluation}
We conduct MOS and SMOS evaluations for different refiners used in our method. Table \ref{tab:refiners} shows the experimental results using  different refiners in MM-TTS including DDPM~\citep{ho2020denoising} and Reflow. According to the results, the Reflow refiner used in MM-TTS achieves better scores in almost every aspect, and the Reflow (1 Euler step) achieves similar results as Reflow (RK45 sampler). This indicates that our proposed Reflow refiner is effective in both quality and sampling speed than DDPM based refiners.

\section{Limitations}
While our approach successfully enables flexible and controllable style transfer using multi-modal prompts, it is important to acknowledge the limitations of our method.
Firstly, the dataset we constructed is relatively small in scale, which restricts its applicability to real-world scenarios. This limitation is evident in the performance comparison on out-of-domain datasets, as illustrated in Table \ref{tab:face_prompt} and \ref{tab:text_prompt}.
Secondly, the text description prompt template we devised has certain limitations, particularly in comprehending more abstract and complex styles. This restricts the system's ability to effectively capture and transfer such nuanced stylistic attributes.
Despite these limitations, we believe that our MM-TTS has taken an important step towards a more universal TTS system with multi-modal capabilities.

\section{Conclusion}
In this work, we propose a general multi-modal prompt based style transfer framework for TTS named MM-TTS.
Specifically, we develop an aligned multi-modal prompt encoder based on CLIP to unify the different modality prompts into the unified style space, empowering the flexible style control by inputting any modality as the prompt during inference stage. In order to transfer the prompt style information to given text content more effectively, we propose SAConv to pay attention to model more local details of style information. Furthermore, we propose a Rectified Flow based refiner to obtain Mel-spectrograms closer to the real domain.
We also construct a multi-modal TTS dataset MEAD-TTS to facilitate future related studies. 
Subjective and  objective experiments demonstrate the superiority of our method. 
\section{Acknowledgments}
This work was supported in part by the National Natural Science Foundation of China under Grants 62276220 and 62001405.

\bibliography{aaai24}

\newpage
\appendix
\onecolumn
\begin{center}{\bf {\LARGE Appendices} }
\end{center}
\begin{center}{\bf {\Large MM-TTS: Multi-Modal Prompt Based Style Transfer for Expressive Text-to-Speech Synthesis} \linebreak}
\end{center}
\section{A~~~MEAD-TTS Dataset} \label{appendix:mead-tts}
\begin{figure*}[!h]
    \centering
    \vspace{-2mm}
    \includegraphics[width=0.77\textwidth]{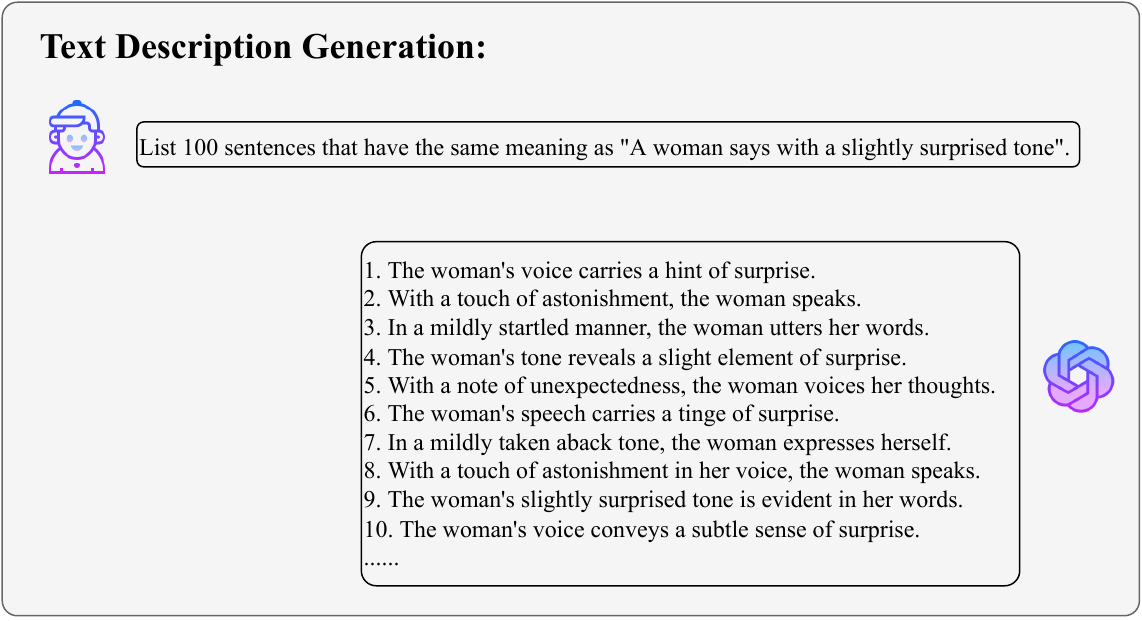}
    \vspace{-2mm}
    \caption{The text description generation process of the text prompts in MEAD-TTS dataset.}
     \vspace{-2mm}
    \label{fig:LLM_prompt}
  \end{figure*}

In this section, we describe the details of MEAD-TTS.
\begin{itemize}
\item \textbf{Overall descriptions}: The MEAD-TTS dataset consists of 31055 pairs of (speech, face image, text description)  data with a total duration of approximately 36 hours for speech modality. There are 47 speakers, including 27 males and 20 females. 
\item \textbf{Text descriptions generation}: We first design a text prompt template like ``A $<$gender$>$ says with a $<$emotion level$>$ $<$emotion$>$ tone'', and then we utilize the GPT-3.5-TURBO to generate the text descriptions having the same meaning as the designed text prompt template. The process is illustrated as Figure \ref{fig:LLM_prompt}.
\end{itemize}

\section{B~~~Data} \label{appendix:data}
In this section, we describe the details of the data usage in training and evaluating MM-TTS.
\begin{itemize}
\item \textbf{Pretraining}: We use LJSpeech~\citep{ljspeech17} as the pretraining data which is a single female speaker dataset and  contains 13100 speech samples with a total duration of approximately 24 hours. We expect to pretrain the model on it to increase the number of transcription data and strengthen the output diversity of TTS.
\item \textbf{Training}: We utilize our constructed MEAD-TTS dataset for the training of MM-TTS, which has data of three modalities to support the training of multi-modal TTS. The details can be seen in Appendix A. 
\item \textbf{Evaluating}: We divide the evaluation into two types: intra-domain evaluation and out-of-domain evaluation. For intra-domain evaluation, we utilize the MEAD-TTS dataset for evaluation of three modalities. For out-of-domain evaluation, we use the test split in LibriTTS~\citep{zen2019} for the evaluation of reference speech based style transfer; we use Oulu-CASIA~\citep{zhao2011facial}, which is a emotional facial image dataset, and LibriTTS transcriptions for the evaluation of face based style transfer; we use the transcriptions of LibriTTS for text description based style transfer.
\end{itemize}

\begin{figure}[!h]
    \centering
    \vspace{-2mm}
    \includegraphics[width=.3\textwidth]{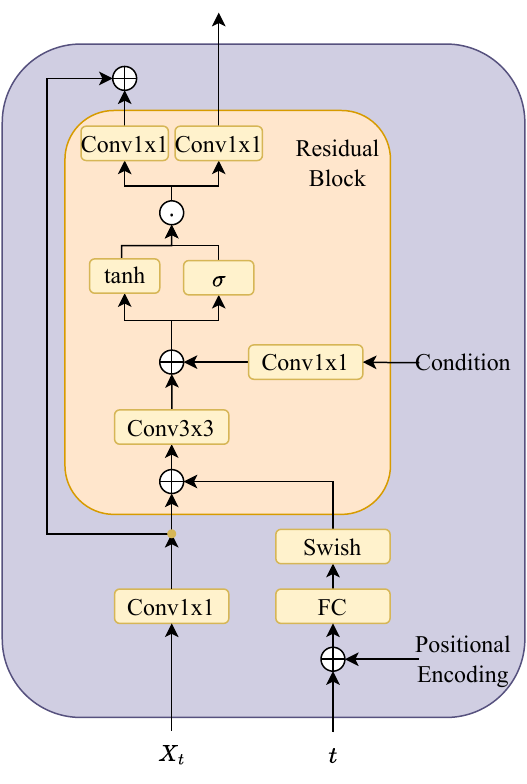}
    \vspace{-2mm}
    \caption{The model architecture of the Reflow Refiner.}
     \vspace{-2mm}
    \label{fig:refiner}
  \end{figure}

\section{C~~~Details of Models}
\subsection{C.1~~~Model Configurations}

We list the model hyper-parameters of MM-TTS (first stage) in Table~\ref{tab:hyperparameters1} and MM-TTS (second stage) in Table~\ref{tab:hyperparameters2}.
\begin{table*}[!h]
    \small
    \centering
    \begin{tabular}{c|c|c}
    \toprule
    \multicolumn{2}{c|}{Hyperparameter}   & MM-TTS (First Stage) \\ 
    \midrule
    \multirow{6}{*}{Aligned Multi-modal Prompt Encoder } 
    &Adapter Linear Layer1 Embed Dim & 256 \\
    &Adapter Linear Layer2 Embed Dim & 3 \\
    &Speech Style Encoder GRU Layers   &   3     \\
    &Speech Style Encoder Conv1D Kernel   &    5     \\
    &Speech Style Encoder Attention Heads   &   1     \\
    &Speech Style Encoder Conv1D Filter Size  &  512  \\

    \midrule
    \multirow{18}{*}{Text-to-Mel Model} 
    &Phoneme Embedding        &    192  \\
    &Encoder Layers           &   4  \\    
    &Encoder Hidden           &  256   \\  
    &Encoder Conv1D Kernel    &   9      \\
	&Encoder Conv1D Filter Size     &  1024   \\    
    &Encoder Attention Heads        &   2 \\ 
    &Encoder Dropout                &  0.1 \\ 
    &Decoder Layers           &   4  \\    
    &Decoder Hidden           &  256   \\  
    &Decoder Conv1D Kernel    &   9      \\
	&Decoder Conv1D Filter Size     &  1024   \\    
    &Decoder Attention Heads        &   2 \\ 
    &Decoder Dropout                &  0.1 \\ 
    &Variance Predictor Conv1D Kernel   & 3  \\
    &Variance Predictor Conv1D Filter Size  & 256  \\
    &Variance Predictor Dropout   & 0.5  \\
    &Kernel Predictor Conv1D Kernel   & 3  \\
    &Kernel Predictor Conv1D Filter Size   & 16  \\
    \midrule

    \multicolumn{2}{c|}{Total Number of Parameters}   &  89.6M \\
    \bottomrule
    \end{tabular}
    \vspace{2mm}
    \caption{Hyperparameters of the first training stage in MM-TTS.}
    \label{tab:hyperparameters1}
    \end{table*}
    
\begin{table*}[!h]
    \small
    \centering
    \begin{tabular}{c|c|c}
    \toprule
    \multicolumn{2}{c|}{Hyperparameter}   & MM-TTS (Second Stage) \\ 
    \midrule
    \multirow{5}{*}{Reflow Refiner} 
    &Reflow Step Embedding   &   256     \\
    &Residual Layers  &  20  \\
    &Residual Channels   &   256     \\
    &WaveNet Conv1D Kernel   &    3     \\
    &WaveNet Conv1D Filter Size   &   512     \\
    \midrule
    \multicolumn{2}{c|}{Total Number of Parameters}   &  15.1M \\
    \bottomrule
    \end{tabular}
    \vspace{2mm}
    \caption{Hyperparameters of the second training stage in MM-TTS.}
    \label{tab:hyperparameters2}
    \end{table*}

\subsection{C.2~~~Reflow Refiner}
Following \citep{jeong2021diff}, we adopt a non-causal WaveNet architecture as our Reflow Refiner. As illustrated in Figure \ref{fig:refiner}, the refiner predicts Gaussian noise from the $t\mbox{-}th$ step noised Mel-spectrogram conditioned on generated over-smoothing Mel-spectrogram and Reflow step embedding. The Reflow step embedding is a sinusoidal embedding with a 128-dimensional encoding vector for each t. The refiner network consists of a stack of 20 residual blocks with Conv1D, tanh, sigmoid, and 1x1 convolutions with 256 residual channels.

\section{D~~~Additional Visualization Results} \label{appendix:results}

\subsection{D.1~~~Reference Speech based Style Transfer}
\begin{figure}[h]
  \centering
  \includegraphics[width=0.81\linewidth]{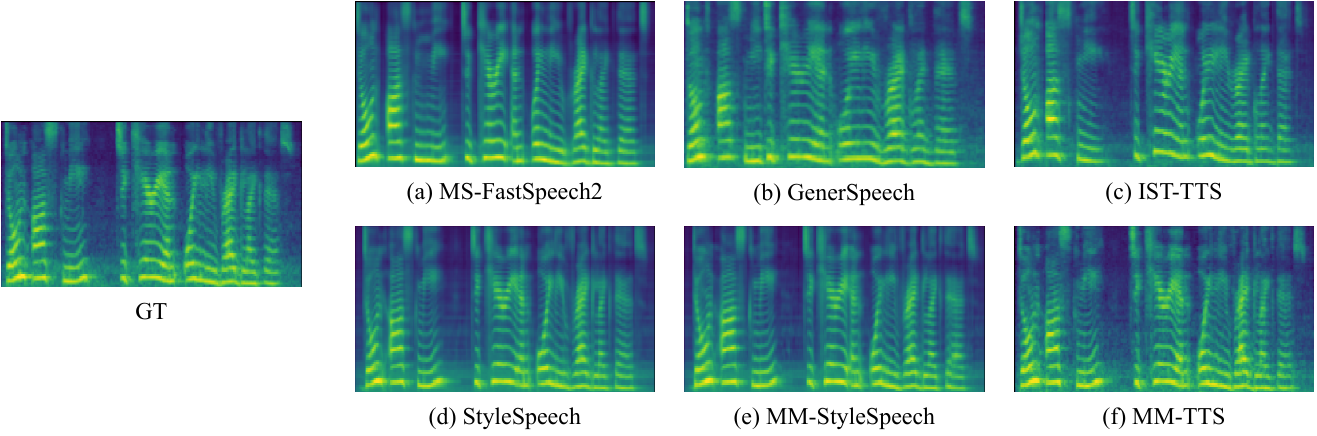}
  \caption{Visualizations of the reference and generated Mel-spectrograms in Parallel style transfer. The content text is ``Don't ask me to carry an oily rag like that.''}
  \label{fig:pst_vis}
\end{figure}

\begin{figure}[h]
  \centering
  \includegraphics[width=0.81\linewidth]{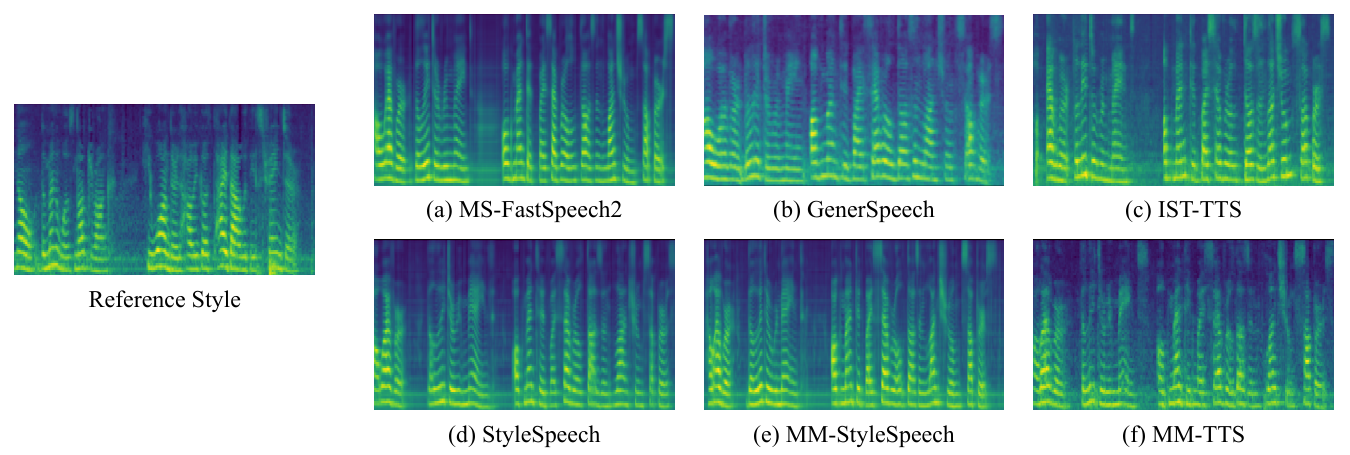}
  \caption{Visualizations of the reference and generated Mel-spectrograms in Non-Parallel style transfer. The reference content text is ``No the man was not drunk he wondered how he got tied up with this stranger.'' and the content text of the generated speech sample is ``However the litter remained augmented by several dozen lunchroom suppers.''}
  \label{fig:npst_vis}
\end{figure}

As shown in Figure \ref{fig:pst_vis} and Figure  \ref{fig:npst_vis}, they are the visualization results of Parallel style transfer and Non-parallel style transfer in reference speech based style transfer respectively.
They demonstrate that our proposed MM-TTS can generate Mel-spectrograms with rich details, which results in more natural voice and precise style transfer.

\subsection{D.2~~~Visualization Results in Ablation Studies} \label{ablation_appendix}
\begin{figure}[t]
  \centering
  \includegraphics[width=0.5\linewidth]{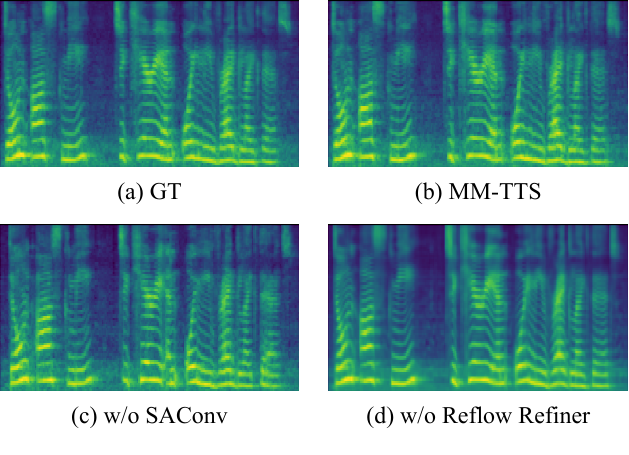}
  \caption{The Mel-spectrogram visualization results of ablation studies.}
  \label{fig:ablation_vis}
\end{figure}

As shown in Figure \ref{fig:ablation_vis}, the visualization results of ablation studies in MM-TTS indicate that the proposed SAConv module significantly improves the style transfer ability and the Reflow Refiner improves the audio quality.

\subsection{D.3~~~Visualization Results in Refiner Evaluations} \label{reflow_appendix}
\begin{figure}[h]
  \centering
  \includegraphics[width=0.66\linewidth]{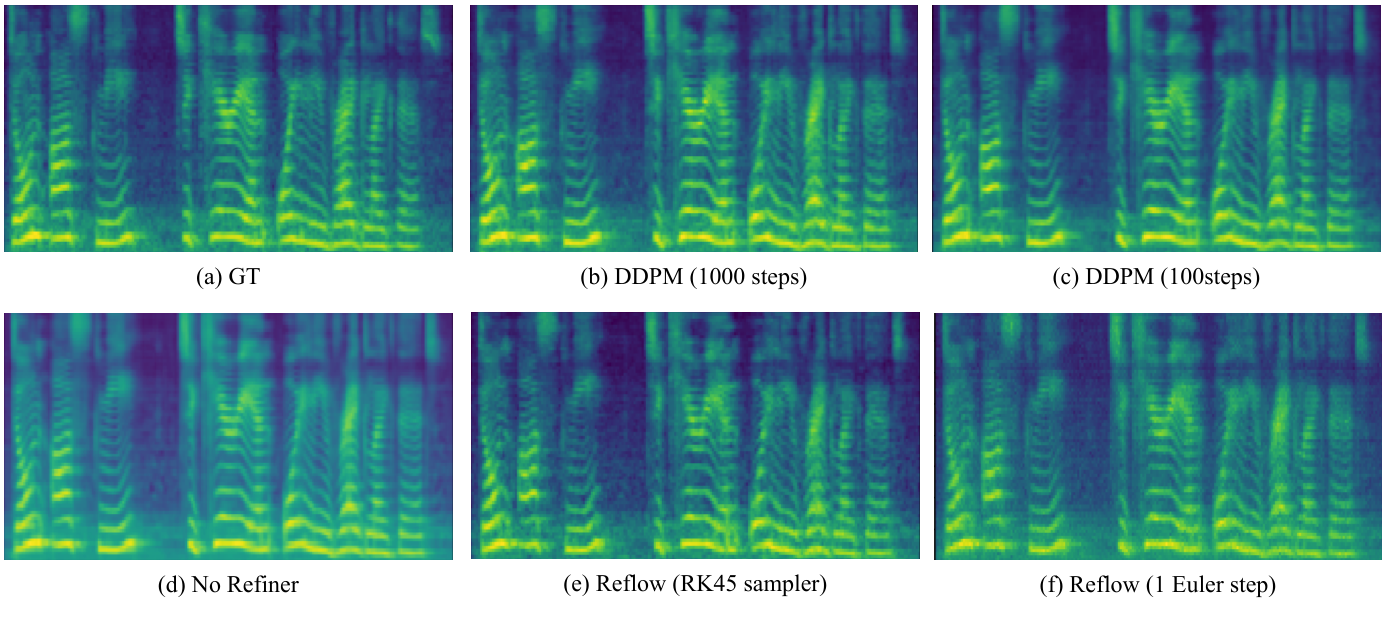}
  \caption{The generated Mel-spectrograms with different refiners.}
  \label{fig:Refiner_figure}
\end{figure}

As shown in Figure \ref{fig:Refiner_figure}, the visualization results of refiner evaluations in MM-TTS demonstrate that our designed Reflow Refiner can achieve similar results with only one step in inference stage.


\end{document}